\documentclass[10pt]{iopart}

\usepackage{iopams, color, enumerate, epsfig, graphicx}
\graphicspath{{graph/}}

\newcommand{\half}{\frac{1}{2}}
\newcommand{\halfpi}{\frac{\pi}{2}}
\newcommand{\lhs}{left-hand side}
\newcommand{\On}{{\rm O}(n)}
\newcommand{\plaq}{{\cal P}}
\newcommand{\Rc}{\check{R}}
\newcommand{\ZN}{\mathbb{Z}_N}

\begin{document}

\title{Discretely Holomorphic Parafermions and Integrable Loop Models}

\author{Yacine Ikhlef$^1$ and John Cardy$^{1,2}$}

\address{$^1$ Rudolph Peierls Centre for Theoretical Physics,
  University of Oxford, 1 Keble Road, Oxford OX1~3NP, United Kingdom}

\address{$^2$ All Souls College, Oxford}

\eads{\mailto{y.ikhlef1@physics.ox.ac.uk}, \mailto{j.cardy1@physics.ox.ac.uk}}

\maketitle

\begin{abstract}
  We define parafermionic observables in various lattice loop models,
  including examples where no Kramers-Wannier duality holds.
  For a particular rhombic embedding of the lattice in the plane
  and a value of the parafermionic spin these variables are
  discretely holomorphic (they satisfy a lattice version of the Cauchy-Riemann equations) 
  as long as the Boltzmann weights satisfy certain linear constraints.
  In the cases considered, the weights then also satisfy the critical Yang-Baxter 
  equations, with the spectral parameter being related linearly to the angle of
  the elementary rhombus.
\end{abstract}

\section{Introduction}
\label{sec:intro}

Holomorphic (and antiholomorphic) fields are the basic building
blocks of conformal field theories (CFTs). They have simple
short-distance expansions among themselves and with other fields
of the theory. Although the prototype of such holomorphic field is
the stress tensor $T(z)$, which is present in all CFTs, many
interesting CFTs contain holomorphic fields with fractional
conformal spin. These are often referred to as parafermionic. The
correlation functions of such holomorphic fields are necessarily
power-behaved, and therefore, in local theories, can exist only in
the massless, conformal case.

Many CFTs are also believed to describe the scaling limit of
critical lattice models. In order to understand the emergence of
holomorphic fields in this limit, the simplest possibility is that
there should exist analogues already in the discrete setting.
These we refer to as {\it discretely holomorphic observables} of the
lattice model. A precise definition is given below.

Recently a number of examples of such observables have been
discovered in well-known lattice models. In models  which possess
a Kramers-Wannier duality symmetry, parafermionic observables can
be defined in terms of suitable products of neighboring order and
disorder variables~\cite{fradkin-kadanoff}. In the case of the
$\ZN$, or clock, models, it was shown in
\cite{rajabpour-cardy} that such parafermions are indeed
discretely holomorphic, but only at the {\it integrable}
critical points identified by Fateev and Zamolodchikov~\cite{FZ}.
In this case the parafermions were defined algebraically directly
in terms of the spin variables and Boltzmann weights of the model.

A second example is the $Q$-state Potts model, which also
possesses a duality symmetry. This model, in its Fortuin-Kasteleyn (FK)
random cluster form, can also be represented in terms of a set of
dense random curves~\cite{baxter-book}. In \cite{riva-cardy}, it
was shown that the parafermions defined in terms of order-disorder
products have support on these curves, and indeed may be
considered as observables depending only on the curves. For
example, the two-point function vanishes unless the two points
happen to lie on the same curve. In \cite{riva-cardy,smirnov} a
simple argument was given that these parafermionic observables are
discretely holomorphic, but only at the critical point of the
Potts model (where it is also known to be integrable).

In this paper we amplify these comments and enlarge the list of
lattice models for which discretely holomorphic observables can be
identified. In particular, we consider various versions of the
$\On$ model which do not possess Kramers-Wannier duality and for
which the definition of a parafermion in terms of an
order-disorder product fails. However, these models all have a
representation in terms of curves on the lattice, and using this
we show that one may still define parafermionic observables,
generalising the construction in the Potts model. Moreover, in all
cases, these turn out to be discretely holomorphic precisely when
the model is integrable and critical.

More generally, we may consider versions of these models where the
weights are anisotropic. In this case, one would expect the
scaling limit to be a rotationally invariant CFT only if the
lattice is embedded in the plane in a particular way. In general,
there is a one-parameter family of such inequivalent embeddings of
a homogeneous lattice. For a given embedding, we find that
discrete holomorphicity holds only for a particular set of
weights. These weights, in all cases, satisfy the Yang-Baxter
equation, with the usual spectral parameter being simply related
to the angle of shear in the embedding.

This is a remarkable result, since the requirement of discrete
holomorphicity, as we will show, gives a set of linear equations
in the weights for a fixed value of the spectral parameter. On the
other hand, the Yang-Baxter equations are cubic functional
relations between the weights at different values of the spectral
parameter.

Let us define more carefully the notion of discrete
holomorphicity. Let $\cal G$ be a planar graph (which will usually
be a regular lattice) embedded in the complex plane, with vertices
at points $\{z_j\}$. Let $F(z)$ be a function defined on the
vertices of the medial lattice: the midpoints $\half (z_i+z_j)$
of the edges $(ij)$ of $\cal G$. Then $F$ is discretely
holomorphic on $\cal G$ if
\begin{equation} \label{eq:holo}
  \sum_{(ij) \in {\plaq}} F \left( (z_i+z_j)/2 \right)\,(z_j-z_i)=0\,,
\end{equation}
where the sum is over the edges of each face, or plaquette,
$\plaq$ of $\cal G$. \Eref{eq:holo} may be thought of as a
discrete version of Cauchy's theorem. However, note that, since
there are fewer equations than unknowns, even in the discrete
setting \eref{eq:holo} is not sufficient to solve a boundary value
problem on $\cal G$. (This should be compared with the example of
the discrete Laplace equation.) In addition, if we try to take the
scaling limit by covering the interior of some domain $\cal D$ in
the plane by a suitable sequence of graphs $\cal G$ whose mesh
size approaches zero, we may deduce that contour integrals vanish,
but Morera's theorem implies that the scaling limit of $F$ is
analytic only if we first assume, or prove, that it is continuous.
(In the Ising case, where the scaling limit has been
proven~\cite{smirnovising}, it was in fact necessary to adopt
another less direct approach.)

In the examples we discuss, the discrete holomorphicity follows
entirely from local properties of the lattice model, and therefore
we may in fact take $F(z)$ to be any correlation function of the
local parafermionic observable with other local fields at other
locations, which may be thought of as a conditional expectation
value. In what follows we shall not distinguish explicitly between
the observable and its correlation functions.

We note that Smirnov~\cite{smirnov} has argued that if one can
find suitable discretely holomorphic parafermionic observables of
curves in lattice models, and also prove that they have a
holomorphic scaling limit satisfying suitable boundary conditions,
such that their dependence on $z$ in simple domains is computable,
then it follows that the scaling limit of the curves is
Schramm-Loewner evolution (SLE$_\kappa$) with a value of $\kappa$
related to the conformal spin. This programme has been carried to
completion for the Ising model, for the curves forming boundaries
of both the spin clusters and the FK clusters~\cite{smirnovising}.

This application of discrete holomorphicity is not the purpose of
the present paper. Rather it is to point out that it holds in a
wider class of models than was observed up to now, and that it
appears to be intimately related to integrability.

However, we should note a general feature of the relation of our
results to SLE. As was first observed in
\cite{friedrich-werner,bauer-bernard}, the statement that a curve
starting on the boundary of a domain is SLE is equivalent in CFT
to the assertion that the boundary operator which inserts the
curve corresponds to a Virasoro highest weight representation with
a null state at level 2, usually denoted by $\phi_{2,1}$ (or
$\phi_{1,2}$). It has conformal weight
$h_{2,1}=(6-\kappa)/(2\kappa)$. Now one of the important
properties of a {\it holomorphic} bulk conformal field $\psi_s(z)$
with spin $s$ is that, when taken to the boundary, it gives a
boundary field with the {\it same} total conformal weight, that is
$s$. If the bulk field is an observable which depends on $N$
curves, which meet locally at the point $z$ but begin at different
points $\{x_j\}$ on the boundary, this implies that the CFT
correlation function
\begin{equation*}
  \langle \psi_s(z) \prod_{j=1}^N \phi_{2,1}(x_j) \rangle
\end{equation*}
is non-zero. By the fusion rules of boundary CFT, this implies
that $\psi_s$ must be a $\phi_{N+1,1}$ field, and therefore that
\begin{equation*}
  s = h_{N+1,1} = \frac{N(2N+4-\kappa)}{2\kappa}\,.
\end{equation*}
Thus if one knows the value of $s$ one may infer the value of
$\kappa$. In this paper we give examples with $N=1,2$, as well as
cases when the curve cannot be simple SLE.

We note that in this paper we shall consider homogeneous,
translationally invariant lattices, but many of the arguments
extend to so-called Baxter lattices, which may be embedded in the
plane isoradially (that is all their faces are rhombi). These have
been discussed with respect to the $\ZN$ models in
\cite{rajabpour-cardy} and the Potts model in \cite{smirnov}.
Discretely holomorphic functions in the Ising model are also
studied in~\cite{mercat}.

The layout of this paper is as follows. In Section~\ref{sec:potts}
we review the arguments for the FK representation of the Potts
model. In Section~\ref{sec:Onsquare} we give a parafermionic
observable for Nienhuis' $\On$ model on the square
lattice~\cite{nienhuis}, and show that it is discretely
holomorphic precisely on the integrable manifolds~\cite{nienhuis}.
In Section~\ref{sec:C21} we extend this to a
model~\cite{warnaar-nienhuis} with two different kinds of loops
which may cross each other. Our arguments in these last two cases
are generalisations of those of Smirnov~\cite{smirnovOn} for the
$\On$ model on the honeycomb lattice.

\section{The self-dual Potts model on the square lattice}
\label{sec:potts}

\begin{figure}
  \begin{center}
    \scalebox{1}{\input{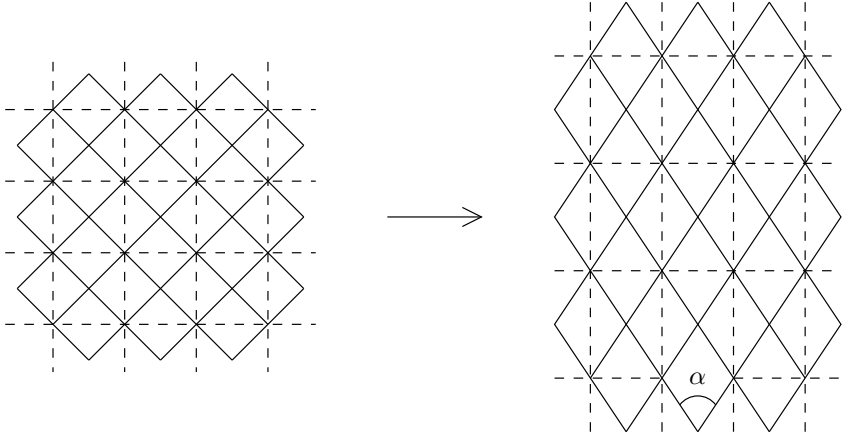}}
    \caption{Left: The original square lattice ${\cal L}$ of the Potts spins (dotted lines)
      and the covering lattice ${\cal M^*}$ (full lines). Right: the deformed lattices, defined
      by the angle $\alpha$.}
    \label{fig:medial}
  \end{center}
\end{figure}

\begin{figure}
  \begin{center}
    \scalebox{1}{\input{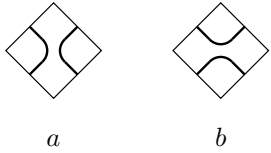}}
    \caption{The elementary plaquette configurations for the loop model, and their Boltzmann weights. Any closed loop has weight $\sqrt{Q}$.}
    \label{fig:tl-bw}
  \end{center}
\end{figure}

The $Q$-state Potts model is a classical spin model on the
lattice. Each vertex carries a spin variable $S_i \in \{1, 2,
\dots, Q\}$, and the Boltzmann weight for a spin configuration is
\begin{equation}
  W \left[ \{S_i\} \right] =
  \prod_{\langle i,j \rangle} \exp \left[ J_{ij} \ \delta(S_i,S_j) \right] \,,
\end{equation}
where the product is on the edges of the lattice, and $J_{ij}$ is
the coupling constant attached to edge~$\langle i,j \rangle$. We
consider the case of the square lattice ${\cal L}$, with coupling
constants $J_1,J_2$ on the horizontal and vertical edges,
respectively.
The model can be reformulated~\cite{baxter-book} as
a dense loop model on the medial lattice ${\cal M}$: this is the
lattice consisting of the midpoints of the original lattice.
The elementary configurations of the
loop model are defined on square plaquettes $\plaq$, which are
the faces of the covering lattice ${\cal M}^*$, the dual of ${\cal M}$ (see {\lhs} of
Figure~\ref{fig:medial}). The
Boltzmann weights of the elementary plaquettes (see
Figure~\ref{fig:tl-bw}) are $(a_r, b_r)$, where
\begin{equation}
  \frac{b_1}{a_1} = \frac{e^{J_1}-1}{\sqrt{Q}}\,,
  \quad \frac{b_2}{a_2} = \frac{\sqrt{Q}}{e^{J_2}-1}\,,
\end{equation}
and $r=1,2$ respectively if the plaquette contains a horizontal or
a vertical edge of the original lattice. Each closed loop carries
a weight $\sqrt{Q}$. Equivalently, the loop model Boltzmann weights can be encoded
in the $\Rc$-matrix $\Rc(u) = a(u) \ 1 + b(u) \ E$, where $E$ (the Temperley-Lieb
generator) is the operator acting on loop diagrams as depicted on the right of
Figure~\ref{fig:tl-bw}, and $u$ is called the spectral parameter.

In this paper, we will restrict ourselves to
the self-dual line:
\begin{equation} \label{eq:sd}
  (e^{J_1}-1)(e^{J_2}-1) = Q\,, \quad J_1,J_2>0\,,
\end{equation}
in the case that the loop model is homogeneous. For $0 \leq Q \leq
4$, the Potts model has a second-order transition on this line. In
this regime, we can write:
\begin{equation}
  \sqrt{Q} = 2 \cos \gamma\,, \quad 0 \leq \gamma \leq \halfpi\,.
\end{equation}

In \cite{riva-cardy,smirnov}, a lattice holomorphic observable
$F_s(z)$ was identified in this loop formulation. The observable
$F_s(z)$ is defined on the midpoints of the plaquette edges as
follows:
\begin{equation} \label{eq:Fs}
  F_s(z) = \sum_{G \in \Gamma(0,z)} P(G) \ e^{-\rmi s \theta(z)}\,,
\end{equation}
where $P(G)$ is the probability of the graph~$G$, $\Gamma(0,z)$ is the
set of loop configurations for which the points $0$ and $z$ belong
to the same loop, and $\theta(z)$ is the winding angle of this
loop from $0$ to $z$. More precisely, we fix arbitrarily a point
$0$ on the edge of a plaquette (this could be in the bulk or on
the boundary of the domain), and a direction for the loop segment
passing through $0$. The winding angle~$\theta(z)$ is defined
incrementally, setting $\theta(0)=0$, and adding $\pm \halfpi$ for
each elementary plaquette on the loop, according to the turn made
by the loop. The real number $s$ is the spin of the parafermion
$F_s(z)$.
In \cite{riva-cardy}, the holomorphicity relation~\eref{eq:holo}
was established
for $F_s$ defined in~\eref{eq:Fs}, at the isotropic self-dual
point: $J_1=J_2= \ln \left( 1 + \sqrt{Q} \right)$. This is only
true if the spin $s$ satisfies:
\begin{equation} \label{eq:s-potts}
  \sqrt{Q} = 2 \sin \frac{\pi s}{2}\,.
\end{equation}
In this section, we adapt these results to the whole (anisotropic) 
self-dual line.

On the one hand, since the loop plaquettes obey a Temperley-Lieb
algebra, the $\Rc$-matrix satisfies the Yang-Baxter equations:
\begin{equation}
  \Rc_{12}(u) \Rc_{23}(v-u) \Rc_{12}(v) = \Rc_{23}(v) \Rc_{12}(v-u) \Rc_{23}(u)
\end{equation}
for the parameterisation $a(u) =\sin u, b(u)=\sin(\gamma-u)$. On the
other hand, let us consider the following {\it deformed model} :
this is the self-dual Potts model with \eref{eq:sd} and anisotropic weights
$J_1 \neq J_2$, defined on the
rectangular lattice with vertical edges scaled by a factor
${\rm cotan} \frac{\alpha}{2}$ (see Figure~\ref{fig:medial}). Various 
arguments (see {\it e.g.} \cite{kadanoff-ceva71,kim-pearce}) support the statement that, in the continuum limit,
the deformed model has the same behaviour as the isotropic self-dual Potts model, if
the angle $\alpha$ is related correctly to the ratio $J_2/J_1$. We
shall obtain the following results for the deformed model:
\begin{enumerate}[(i)]
  \item $F_s$ is holomorphic on the lattice if the spin $s$ satisfies the 
    relation~\eref{eq:s-potts}.
  \item There exists a linear relation between the spectral parameter $u$ 
    and the angle $\alpha$.
\end{enumerate}

\begin{figure}
  \begin{center}
    \scalebox{1}{\input{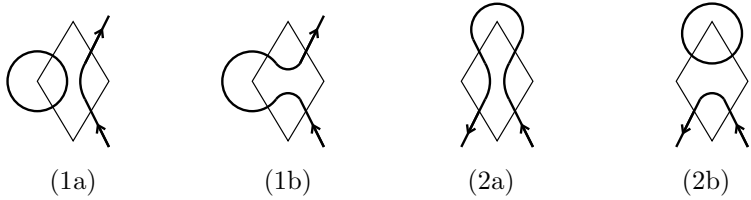}}
    \caption{Loop configurations with a loop connecting $0$ to the elementary plaquette $\plaq$.}
    \label{fig:holo-potts}
  \end{center}
\end{figure}

Let us describe in more details the deformed geometry: the
plaquettes become rhombi with internal angles $\alpha,
\pi-\alpha$, and the increments of $\theta(r)$ are now $\pm
\alpha, \pm(\pi-\alpha)$ on each plaquette. The {\lhs} of
\eref{eq:holo} has only contributions from the loop configurations
in which a closed loop $C$ connects $0$ to two edges of the
plaquette~$\plaq$. Let $G$ be one such configuration, where the
plaquette~$\plaq$ is in configuration (a) of
Figure~\ref{fig:tl-bw}, and $G'$ the configuration differing from
$G$ only by the configuration of the plaquette~$\plaq$. Our method
is to show that the contributions to the {\lhs} of \eref{eq:holo}
from $G,G'$ cancel each other. There are essentially two
inequivalent external connectivities of the loop outside of
$\plaq$, as shown in Figure~\ref{fig:holo-potts}. In the case (1),
the probabilities of $G,G'$ are related by $P(G')/b = P(G)/(a
\sqrt{Q})$, whereas in case (2), one has $P(G')/(b\sqrt{Q})=
P(G)/a$. The requirement of cancellation of contributions from
$G,G'$ yields a linear system for the Boltzmann weights $a,b$:
\begin{equation} \label{eq:syst-potts}
  \left\{
    \ \eqalign{
      (1+\mu)\sqrt Q \ a + (1+\mu-\lambda-\mu \lambda^{-1}) \ b = 0 \\
      (1+\mu-\lambda^{-1}-\mu \lambda^{-1}) \ a + (1-\mu\lambda^{-1}) \sqrt Q \ b = 0\,,
    }
  \right.
\end{equation}
where we have set $\lambda= e^{\rmi \pi s}\,, \mu = e^{\rmi \alpha(s+1)}$.
The determinant of the system reads:
\begin{equation}
  \lambda^{-1} (1+\mu) (1-\mu \lambda^{-1}) \left[ \lambda^2 +(Q-2) \lambda +1 \right]\,.
\end{equation}
This has to vanish for the system to have a non-trivial solution.
The last factor vanishes for $\lambda = - e^{\pm 2 \rmi \gamma}$. We
choose the minus sign, corresponding to the exponent:
\begin{equation} \label{eq:s-potts2}
  s= 1 - \frac{2\gamma}{\pi}\,.
\end{equation}
Note that this value satisfies the condition~\eref{eq:s-potts}. The relative Boltzmann weight is then:
\begin{equation}
  \frac{b}{a} = - \frac{\cos \frac{(s+1)\alpha}{2}}
  {\cos \left[ \gamma + \frac{(s+1)\alpha}{2} \right]}\,.
\end{equation}
Comparing with the parameterisation $a(u), b(u)$ for the $\Rc$-matrix, we get:
\begin{equation} \label{eq:u-alpha}
  \frac{(s+1)\alpha}{2} = \halfpi - u\,.
\end{equation}
Therefore, we recover the result of~\cite{kim-pearce}, that the angle $\alpha$ is linearly 
related to the spectral parameter $u$. Note that~\eref{eq:u-alpha} is different from the usual
relation $\alpha = \pi u /\gamma$. This can be corrected by setting the loop's angle increments
on a plaquette to $\beta, \pi-\beta$. Then the system~\eref{eq:syst-potts} still holds, if we set
$\mu = e^{\rmi (\alpha+ \beta s)}$. The angle $\beta$ can then be tuned to recover the appropriate relation
between $\alpha$ and $u$.

The scaling limit of the self-dual Potts model is described by a
Coulomb gas with central charge and conformal weights given by:
\begin{equation} \label{eq:cg}
  \eqalign{
    c = 1 - \frac{6(1-g)^2}{g} \\
    h_{r,r'} = \frac{(gr-r')^2 - (1-g)^2}{4g}\,,
  }
\end{equation}
where the coupling constant $g$ is given by $g =
1-\frac{\gamma}{\pi}$. One can then write \eref{eq:s-potts2} as
$s=h_{3,1}$. According to our earlier observation, this is
consistent with the fact that there are \em two \em curves meeting
at the observation point $z$.

\section{The $\On$ loop model on the square
lattice}\label{sec:Onsquare}

\begin{figure}
  \begin{center}
    \scalebox{1}{\input{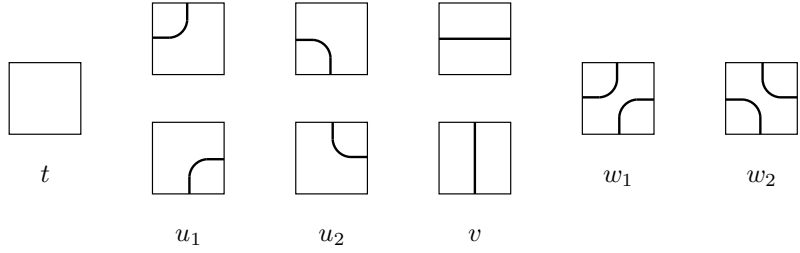}}
    \caption{Vertices of the $\On$ loop model on the square lattice.
      Each closed loop has a weight $n$.}
    \label{fig:On-bw}
  \end{center}
\end{figure}

In this section, we find solutions of the holomorphicity
equations~\eref{eq:holo} for the $\On$ loop model on the square
lattice~\cite{nienhuis}. This is a dilute loop model, defined by
the vertices in Figure~\ref{fig:On-bw}, where each closed loop
carries a weight given by:
\begin{equation*}
  n = -2 \cos 2 \eta\,, \quad 0 \leq \eta \leq \halfpi\,.
\end{equation*}
Note that we have grouped the anisotropic weights so that there is
symmetry under reflections in the diagonal axes. This symmetry is
preserved when the plaquettes are deformed into rhombi.
Since every loop configuration has an even number of plaquettes of
type $u_1$ or $u_2$, the change $(u_1,u_2) \to (-u_1,-u_2)$ does not
affect the Boltzmann weights.

\begin{figure}
  \begin{center}
    \includegraphics[scale=1]{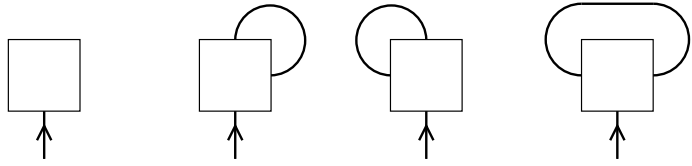}
    \caption{Loop configurations with one edge of the plaquette~$\plaq$
      connected to point $0$.}
    \label{fig:On-holo}
  \end{center}
\end{figure}

We consider the observable~$F_s(z)$, defined similarly to
\eref{eq:Fs}, except that the sum is now over graphs where there
is an {\it oriented open path} going from $0$ to $z$. A similar
observable has been considered by Smirnov~\cite{smirnovOn} for the
case of the hexagonal lattice. Now consider the contributions to
the holomorphicity equation~\eref{eq:holo} where the first time the 
oriented curve, starting at $0$, enters the chosen plaquette $\plaq$ 
is through a particular edge, for example
the lowermost in Figure~\ref{fig:On-holo}). There are four
inequivalent external connectivities as shown. We imagine these to
be fixed, summing over all internal configurations consistent with
them. This yields the linear system for the Boltzmann weights:
\numparts
\begin{eqnarray}
  t + \mu u_1 - \mu \lambda^{-1} u_2 - v
  = 0 \label{eq:syst-On1} \\
  -\lambda^{-1} u_1 + n u_2 + \lambda \mu v - \mu \lambda^{-1} (w_1+n w_2)
  = 0 \label{eq:syst-On2} \\
  n u_1 - \lambda u_2 - \mu \lambda^{-2} v + \mu (n w_1 + w_2)
  = 0 \label{eq:syst-On3} \\
  - \mu \lambda^{-2} u_1 + \mu \lambda u_2 + n v - \lambda^{-2} w_1 - \lambda^2 w_2
  = 0\,, \label{eq:syst-On4}
\end{eqnarray}
\endnumparts
where we have set: $\lambda = e^{\rmi \pi s}\,, \varphi = (s+1) \alpha\,, \mu = e^{\rmi \varphi}$. For real Boltzmann weights, (\ref{eq:syst-On1}--\ref{eq:syst-On4}) 
are four complex linear equations for six real unknowns $(t,u_1,u_2,v,w_1,w_2)$, 
and we have the relations:
\begin{eqnarray*}
  \fl {\rm Im}\ \left[
    (n+1) \ \eref{eq:syst-On1} - \lambda \mu^{-1} \ \eref{eq:syst-On2}
    + \mu^{-1} \ \eref{eq:syst-On3}
  \right] = 0 \\
  \fl {\rm Im}\ \left[
    \lambda \mu^{-1} (\lambda^2-n \lambda^{-2}) \ \eref{eq:syst-On2}
    + \mu^{-1} (n \lambda^2 - \lambda^{-2}) \ \eref{eq:syst-On3}
    - (n^2-1) \ \eref{eq:syst-On4}
  \right] = 0\,.
\end{eqnarray*}
Thus, we can generally reduce (\ref{eq:syst-On1}--\ref{eq:syst-On4}) to a $6 \times 6$ real system.

\begin{figure}
  \begin{center}
    \input{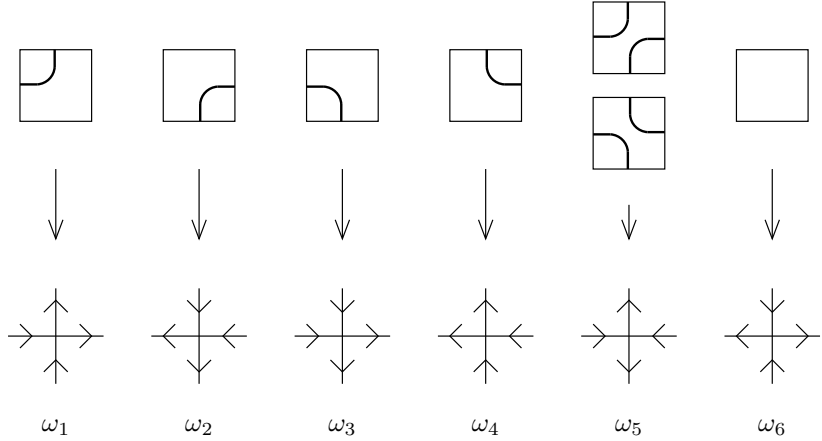}
    \caption{Mapping of the $\On$ model onto the six-vertex model for $n=1,v=0$.}
    \label{fig:On-6v}
  \end{center}
\end{figure}

There are two classes of solutions, for vanishing and non-vanishing $v$.
First, if $v=0$, then the configurations corresponding
to \eref{eq:syst-On4} never occur, and so this equation does not hold.
In the special case $n=1$, there exists a non-trival solution 
{\it for any value of $s$}:
\begin{equation}
  t= \sin \pi s\,, \ u_1=\sin(\varphi-\pi s)\,, \ u_2=\sin \varphi\,, 
  \ w_1+w_2=\sin \pi s\,.
\end{equation}
This model can be mapped onto the six-vertex model (see
Figure~\ref{fig:On-6v}), with weights
$\omega_1=\omega_2=\sin(\varphi-\pi s), \omega_3=\omega_4=\sin
\varphi, \omega_5=\omega_6=\sin \pi s$. The corresponding
anisotropy parameter~\cite{baxter-book} is $\Delta = \cos \pi s$.
This is an example of a model admitting a holomorphic observable
on the lattice, but for which the scaling limit of the
corresponding curve cannot be described by simple SLE. This is
because, for ordinary SLE, the central charge of the CFT is
directly related to the SLE parameter $\kappa$
\cite{friedrich-werner, bauer-bernard} and hence to the conformal
spin $s$: $c = 2s(5-8s)/(2s+1)$. In the present case, since the
boundary conditions for the six-vertex model are not twisted, its
scaling limit has central charge $c=1$ for all $\Delta$. However
the conformal spin $s$ varies continuously with $\Delta$.
Therefore the scaling limit of the curve can be SLE, with
$\kappa=4$, for at most one value (in fact $\Delta=1/\sqrt2$.) We
conjecture that other values of $\Delta$ correspond to
SLE$(4,\rho)$.

For $v=0\,, n \neq -1$, we get a $5 \times 5$ linear system, with determinant
$(n^2-1)^2 \sin \varphi \ \sin (\varphi-\pi s)$. 
Imposing $\sin \varphi=0$ yields $(s+1) \alpha = m \pi$ and in turn $s=m'$, where $m,m'$ are integers.
Thus, for the solution to exist at any value of $\alpha$, we have to set $s=-1$.
The Boltzmann weights are then:
\begin{equation} \label{eq:n+1}
  t = -u_1-u_2\,, \ w_1= -u_1\,, \ w_2= -u_2\,.
\end{equation}
The solution of the case $\sin (\varphi-\pi s)=0$ is similar, and leads to the same Boltzmann weights 
and spin $s=-1$.
If we change the sign of $u_1,u_2$, then the model~\eref{eq:n+1}
is equivalent to the dense loop model of Section~\ref{sec:potts}, with parameters
$\sqrt{Q}=n+1, a=u_1, b=u_2$. To see this, fill empty spaces with loops of weight
$1$ ({\it ghost loops}). The local weights do not depend on the type of loops
involved (actual or ghost loops), so each loop has an overall weight $n+1$.
As a consequence, the dense loop model has a lattice antiholomorphic observable ($s<0$),
besides the holomorphic one found in~\cite{riva-cardy}. However, several arguments rule out the 
hypothesis that $F_{s=-1}(z)$ corresponds to an antiholomorphic field in the continuum limit.
First, $F_{s=-1}(z)$ is lattice antiholomorphic for any $Q>0$, whereas it is well known that the
self-dual Potts model is only critical for $0 \leq Q \leq 4$. Furthermore, the ratio $u_1/u_2$ in 
\eref{eq:n+1} does not depend on the angle $\alpha$, which means that the same model has an 
antiholomorphic observable for any deformation angle: this is not acceptable physically in the 
continuum limit. So we conclude that, in the case $v=0$ and generic $n \neq -1$, the holomorphicity 
conditions~\eref{eq:holo} for the dilute $\On$ model merely lead to the
case of the dense loop model, but the corresponding $F_s(z)$ is not a candidate for an
antiholomorphic field in the continuum limit.

Let us now discuss the solutions of second class ($v\neq 0$), for a generic value of $n$.
We get the $6 \times 6$ real system:
\begin{equation*}
  \{{\rm Re} \ \eref{eq:syst-On1}, {\rm Re} \ \eref{eq:syst-On2}, {\rm Im} \ \eref{eq:syst-On2},
  {\rm Re} \ \eref{eq:syst-On3}, {\rm Im} \ \eref{eq:syst-On3}, {\rm Re} \ \eref{eq:syst-On4} \}\,,
\end{equation*}
with determinant: $(n^2-1) \sin \varphi \sin (\varphi-\pi s) \left(2 \cos 4\pi s -3n+n^3 \right)$.
Non-trival solutions exist if the spin satisfies:
\begin{equation} \label{eq:s-On1}
  \cos 4 \pi s = \cos 6 \eta\,.
\end{equation}
The various solutions to \eref{eq:s-On1} can be parameterised by
extending the range of $\eta$ to $[-\pi,\pi]$, and setting:
\begin{equation} \label{eq:s-On2}
  s = \frac{3 \eta}{2\pi} - \half\,.
\end{equation}
Then, we get the second class of
solutions, with Boltzmann weights:
\begin{equation} \label{eq:bw-On}
  \eqalign{
    t   = -\sin \left( 2\varphi-{3\eta}/{2} \right)
      + \sin {5\eta}/{2} - \sin {3\eta}/{2} + \sin {\eta}/{2} \\
    u_1 = -2 \sin \eta \cos \left( {3\eta}/{2} - \varphi \right) \\
    u_2 = -2 \sin \eta \sin \varphi \\
    v   = -2 \sin \varphi \cos \left( {3\eta}/{2} - \varphi \right) \\
    w_1 = -2 \sin(\varphi-\eta) \cos \left( {3\eta}/{2} - \varphi \right) \\
    w_2 = 2 \cos \left({\eta}/{2} - \varphi \right) \sin \varphi\,.
  }
\end{equation}
A remarkable fact is that the weights~\eref{eq:bw-On} are a solution of the
Yang-Baxter equations for the $O(n)$ loop model on the square
lattice. Indeed, after a change of variables $\varphi \to \psi+
(\pi+\eta)/4$, \eref{eq:bw-On} coincides with
the integrable weights in~\cite{nienhuis}. So, by solving the
holomorphicity equations~(\ref{eq:syst-On1}--\ref{eq:syst-On4})
on a deformed lattice, we recovered the integrable weights.

Now we interpret our findings in terms of CFT. First, in the
regime $0 \leq \eta \leq \pi$ (branches 1 and 2 in
\cite{blote-nienhuis}), on the basis of numerical diagonalisation
of the transfer-matrix~\cite{blote-nienhuis} and analysis of the
Bethe Ansatz equations \cite{batchelor-etal, warnaar-etal}, it has
been argued that the model~\eref{eq:bw-On} has the same continuum
limit as the $\On$ model on the hexagonal
lattice~\cite{nienhuis82}: it is described by a Coulomb gas of
coupling constant $g=2\eta/\pi$, with central charge and critical
exponents given by \eref{eq:cg}. The value of the spin can then be
written as $s=h_{2,1}$. In the regime $-\pi \leq \eta \leq 0$, the
continuum limit is no longer described by a simple Coulomb gas.
Rather, it has central charge:
\begin{equation} \label{eq:c34}
  c = \frac{3}{2} - \frac{6(1-g')^2}{g'}\,,
\end{equation}
where $g'=2(\pi+\eta)/\pi$. The relation between
$s$~\eref{eq:s-On2} and $c$~\eref{eq:c34}, which is $c =
(2-5s-16s^2)/(2s+2)$, is not consistent with SLE, so this is
another example of a model with a holomorphic observable whose
curves are not described by simple SLE. Note that in this case $c$
may be greater than one, so some variant of SLE such as that considered by
Bettelheim {\it et al.}~\cite{bettelheim} may be the correct
description.

\section{The $C_2^{(1)}$ loop model}
\label{sec:C21}

\begin{figure}
  \begin{center}
    \scalebox{1}{\input{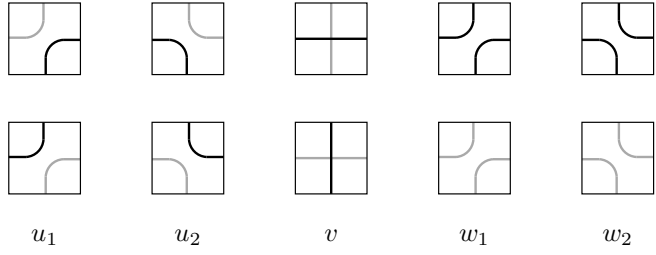}}
    \caption{Vertices of the $C_2^{(1)}$ loop model on the square lattice.
      Each closed loop has a weight $n$.}
    \label{fig:C21-bw}
  \end{center}
\end{figure}

A simple generalisation of the loop model of
Section~\ref{sec:potts} to a model with two loop colours was
introduced in~\cite{warnaar-nienhuis}. It is a dense loop model on
the square lattice, where each loop can be either black or grey,
and carries a weight $n$ given by:
\begin{equation*}
  n = -2 \cos 2 \eta\,, \quad 0 \leq \eta \leq \halfpi\,.
\end{equation*}
The vertices of the model are shown in Figure~\ref{fig:C21-bw}.
For fixed $n$, the model has a one-parameter family of Boltzmann
weights that satisfy the Yang-Baxter equations~\cite{warnaar-nienhuis}.

To describe the observable $F_s(z)$ that will be shown to be
holomorphic, we first introduce a special kind of defect in the
loop model: this defect sits on the midpoint of a plaquette edge,
and consists in a change of colour for the loop passing through.
The observable~$F_s(z)$ is then defined similarly
to \eref{eq:Fs}, except that the sum is now over graphs~$G$ where
$0$ and $z$ belong to the same loop, and each of these points
carries a defect. The angle~$\theta(z)$ is defined as the sum of
the windings of each loop strand from $0$ to $z$. For a given
plaquette~$\plaq$ in \eref{eq:holo}, the graphs that
contribute are those for which two edges of the plaquette~$\plaq$
are connected to the point $0$, and one edge contains a defect.
The holomorphicity conditions~\eref{eq:holo} yield the system:
\numparts
\begin{eqnarray}
  n u_1 - \lambda u_2 - \mu \lambda^{-1} v + \mu (n w_1 + w_2) = 0 \label{eq:C21-syst1} \\
  -\lambda^{-1} u_1 + n u_2 + \mu v - \mu \lambda^{-1} (w_1 + n w_2) = 0 \label{eq:C21-syst2} \\
  -\mu \lambda^{-1} u_1 + \mu u_2 + n v - \lambda^{-1} w_1 - \lambda w_2 = 0\,, \label{eq:C21-syst3}
\end{eqnarray}
\endnumparts
where we have set: $\lambda = e^{2 \rmi \pi s}\,, \varphi=(2s+1)\alpha\,, \mu=e^{\rmi \varphi}$. For real Boltzmann weights, we have the linear relation:
\begin{equation*}
  \fl {\rm Im}\ \left[ \mu^{-1}(n\lambda-\lambda^{-1}) \ \eref{eq:C21-syst1}
    +\lambda \mu^{-1}(\lambda-n\lambda^{-1}) \ \eref{eq:C21-syst2}
    +(n^2-1) \ \eref{eq:C21-syst3}
  \right] = 0\,.
\end{equation*}
Thus, for $n^2 \neq 1$, we can ignore ${\rm Im} \ \eref{eq:C21-syst3}$, and the remaining $5 \times 5$ real system has determinant: $(n^2-1) \sin \varphi \sin(\varphi-2\pi s) \left( 2 \cos 4\pi s -3n +n^3 \right)$.
It has a non-trivial solution for:
\begin{equation}
  \cos 4 \pi s = \cos 6 \eta\,.
\end{equation}
There are two branches of solutions, which can be obtained by
setting $s = (3\eta -\pi)/(2\pi)$, and extending the range of
$\eta$ to $[0, \pi]$. The Boltzmann weights then read:
\begin{equation} \label{C21-bw}
  \eqalign{
    u_1 = \sin \eta \sin(\varphi-3\eta) \\
    u_2 = -\sin \eta \sin \varphi \\
    v   = -\sin \varphi \sin(\varphi-3\eta) \\
    w_1 = -\sin(\varphi-\eta) \sin(\varphi-3\eta) \\
    w_2 = -\sin(\varphi-2\eta) \sin \varphi\,.
  }
\end{equation}
These are exactly the integrable weights given in \cite{warnaar-nienhuis}, with the change
of variables $(\eta \to \lambda, \varphi \to u)$. So the $C_2^{(1)}$ loop model is another
example where the solution of the lattice holomorphicity equations~\eref{eq:holo} satisfies
also the Yang-Baxter equations.

\section{Summary}
In this paper we have identified several more examples of
discretely holomorphic parafermionic observables in lattice
models, and shown that this requirement always appears to pick out
the critical points which are also integrable in the sense of
Yang-Baxter. It would of course be important to have a more
general understanding of this. It may give a simpler route to
finding new integrable models. These parafermions are natural
discrete candidates for parafermionic holomorphic conformal fields
in the scaling limit, and thus can also suggest new structures in
particular CFTs. In some cases the models we have considered
correspond to CFTs with $c\geq1$, so the scaling limit of the
corresponding curves is probably described by some modification of
simple SLE. The loop models we have considered may also be mapped
to generalised restricted solid-on-solid (RSOS) models, and, from
this point of view, it would be interesting to understand the
coset construction of the corresponding CFT and its relation to
the parafermionic fields.

\ack{We thank Paul Fendley and Stas Smirnov for important discussions.
This work was supported by EPSRC Grant EP/D050952/1.}

\end{document}